\documentclass[11pt]{article}
\usepackage{moriond,epsfig}

\def\Journal#1#2#3#4{{#1} {\bf #2}, #3 (#4)}

\def\APJ{\em Astrophys. J.}
\def\APJL{\em Astrophys. J. Lett}

\def\be{\begin{equation}}
\def\ee{\end{equation}}
\def\bea{\begin{eqnarray}}
\def\eea{\end{eqnarray}}

\include{moriond.sty}

\begin{document}

\begin{center}
To appear in {\it Proceedings of XXI Moriond Conference: Galaxy Clusters and the High Redshift Universe Observed in X-rays}, edited by D. Neumann, F. Durret, \& J. Tran Thanh Van, in press
\end{center}

\vspace*{4cm}
\title{CHANDRA OBSERVATIONS OF A CLUSTER AND TWO QUASARS AT HIGH REDSHIFT}

\author{C.R. CANIZARES, T.E. JELTEMA,  T. FANG, M.W. BAUTZ,}
\author{M.R. MALM, H.L. MARSHALL, G. BRYAN}

\address{Center for Space Research and Department of Physics\\
Massachusetts Institute of Technology, Cambridge, MA 02139}

\author{M. DONAHUE}

\address{Space Telescope Science Institute, Baltimore, MD 21218}

\author{G.P. Garmire}

\address{The Pennsylvania State University, University Park, PA 16802}

\maketitle\abstracts{
We observed MS~1054$-$0321, the highest redshift cluster of galaxies in the 
{\itshape Einstein\/} Medium Sensitivity Survey (EMSS), with the {\itshape 
Chandra\/} ACIS-S detector.  We find the X-ray temperature of the cluster to 
be $10.4^{+1.7}_{-1.5}$ keV, slightly lower than that inferred previously
and in agreement with the observed 
velocity dispersion and that found from weak lensing.  We confirm 
significant substructure in the form of two distinct clumps in the 
X-ray distribution.  The eastern clump seems to coincide with the 
main cluster component and has a temperature of $10.5^{+3.4}_{-2.1}$ keV, 
approximately the same as the average spectral temperature for the 
whole cluster.  The western clump is cooler, with a temperature 
of $6.7^{+1.7}_{-1.2}$ and may be a subgroup falling into the cluster.  
Though the presence of substructure indicates that this cluster is not 
fully relaxed, cluster simulations suggest that we will underestimate 
the mass, and we can, therefore, use the mass to constrain $\Omega_m$.  
From the overall cluster X-ray temperature we find the virial mass of 
the cluster to be at least $4.5 \times 10^{14} h^{-1} M_{\odot}$.  
We revisit the cosmological implications of the existence of such a 
hot, massive cluster at a relatively early epoch.  Despite the lower 
temperature, we still find that the existence of this cluster constrains 
$\Omega_m$ to be less than one.  If $\Omega_m = 1$ and assuming Gaussian 
initial 
perturbations, we find the probability of observing MS~1054 in the EMSS 
is $\sim 7 \times 10^{-4}$.
We also present the first high-resolution X-ray spectra of two high-redshift quasars, S5 0836+710 and PKS 2149-306, obtained with the {\itshape Chandra} High Energy Transmission Grating Spectrometer (HETGS). The primary goal of this observation is to use the high spectral resolving power of the HETGS to detect X-ray absorption produced by a hot intergalactic medium. The continuum of both quasars can be fitted by absorbed power laws. Excess continuum absorption above the Galactic value is found in S5 0836+710, as evidenced in {\sl ASCA} and {\sl ROSAT} observations. No significant emission or absorption feature is detected in either source at $\pm3\sigma$ level. Based on the detection limits we constrain the properties of possible emitters and absorbers. Absorbers with a column density higher than $8\times10^{16}\ cm^{-2}$ for O VIII or $5\times10^{16}\ cm^{-2}$ for Si XIV would have been detected. We propose a method to constrain the cosmological parameters (namely $\Omega_{0}$ and $\Omega_{b}$) via the X-ray forest theory, but current data do not give significant constraints.  We also find that combined with the constraints from the distortion of the CMB spectrum, the X-ray Gunn-Peterson test can marginally constrain a uniform, enriched IGM.
}

\section{Introduction}

This paper summarizes two {\itshape Chandra} X-ray Observatory
 observations of objects at 
high redshift.
MS1054$-$0321 is the highest redshift cluster in the Einstein Extended 
Medium Sensitivity Survey (EMSS) with z = 0.83, whose very existence has 
cosmological implications. Imaging observations with ACIS are used to 
measure its temperature, substructure and estimate its mass. The full
report of this work is given in Jeltema {\it et al}.~\cite{J01}. Two high z quasars were observed 
with the High Energy Transmission Grating Spectrometer. Here the primary goal
was to search for absorption lines from intervening material in the
putative hot IGM. Such a medium is thought to be the primary
reservoir for Baryons in the present epoch. A full report of this
observation is given in Fang {\it et al}.~\cite{F01}.

\section{MS1054$-$0321}

  This cluster has 
previously been observed at X-ray wavelengths with {\itshape ASCA\/} 
and {\itshape ROSAT\/}.  Its mass has been estimated from the {\itshape 
ASCA\/} X-ray temperature of $12.3^{+3.1}_{-2.2}$ (Donahue {\it et al}.~\cite{D98}), from 
$\beta$-model fits to the {\itshape ROSAT/HRI\/} data (Neumann \& Arnaud~\cite{NA00}), 
and, at optical wavelengths, from its weak lensing signal (Hoekstra, Franx \& Kuijken~\cite{HFK00}; 
Luppino \& Kaiser~\cite{LK97}). All of these methods indicate that MS~1054 is a massive 
cluster, which in conjunction with its high redshift implies 
$\Omega_m < 1$.  Substructure, an indication that the cluster 
may not be relaxed, was seen in both the {\itshape ROSAT/HRI\/} 
observation and the weak lensing data. See also the contributions 
of M. Donahue and P. Marshall to these proceedings.

We use $H_0 = 100 h$ km s$^{-1}$ Mpc$^{-1}$.  For $q_0 = 0.5$ and $\Lambda = 0$, one arcminute is 249 $h^{-1}$ kpc at the cluster's redshift.

{\itshape Chandra\/} observed MS~1054 with the back-illuminated ACIS-S3 detector on 2000 April 21-22 for 91 ks. Twenty-three point sources were identified and removed from the data. The background was estimated from the local background on the ACIS-S3 chip.  The net cluster count rate was 0.13 counts/sec in a $2^{\prime}$ radius region and the 0.3-7.0 keV band. 

The overall shape of the X-ray emission is elongated, and the existence of substructure is evident as two distinct peaks can be seen in the X-ray contours separated by about $1.2^{\prime}$ (300 $h^{-1}$ kpc). These general features were already noted by Donahue {\it et al}.~\cite{D98} and Neumann \& Arnaud~\cite{NA00}, but are seen more clearly here.

Spectra were analyzed using data in the 0.8-7.0 keV band using XSPEC with standard models. Foreground absorption was fixed at the galactic value of $3.6 \times 10^{20}$ atoms/cm$^2$, the redshift was fixed at 0.83, and the iron abundance and temperature were free to vary.  The best-fit temperature is $10.4^{+1.7}_{-1.5}$ keV with an abundance of $0.26\pm0.15$ and a luminosity in the 2-10 keV band of $6.3 \times 10^{44}$ h$^{-2}$ ergs s$^{-1}$ ($q_0 = 0.1$).  All uncertainties are 90\% confidence levels.  The detection of the iron emission line allows us to fit for the redshift, which gives $z = 0.83\pm0.03$.  The fit is good with a reduced $\chi^2$ of 1.03 for 239 degrees of freedom.  

The best fit cluster temperature from the {\itshape Chandra\/} data is somewhat lower than the {\itshape ASCA\/} temperature of $12.3^{+3.1}_{-2.2}$ keV (Donahue {\it et al}.~\cite{D98}).  However, the two results agree within the 90\% limits, and there is no statistically significant discrepancy. 

We investigated the temperature of each of the two primary sub-features, taking spectra in a 0.41$^{\prime}$ radius circle around the brightest point in each clump.  For the eastern clump, the best fit temperature is $10.5^{+3.4}_{-2.1}$ keV with a reduced $\chi^2$ of 0.94 for 75 degrees of freedom.  This temperature agrees well with the overall cluster temperature.  For the western clump, the temperature is $6.7^{+1.7}_{-1.2}$ with a reduced $\chi^2$ of 1.08 for 62 degrees of freedom.  The eastern clump is somewhat hotter than the western clump, with the lower limit of the 90\% confidence range for the eastern clump just equaling the upper limit for the western clump.

To get an indication of temperature variations in the cluster, we created a map of the hardness ratio. The primary result of this analysis is that the interclump region is cooler than the cluster as a whole, indicating the
absence of a shock between the two sub-clumps.  

The western clump could either be a subgroup falling into the cluster or a foreground group of galaxies.  Since we detect an iron line, we fit for the redshift of the western clump in XSPEC and get a value of z = $0.84\pm0.02$, which agrees well with the redshift of the cluster.  Substructure can also be seen in the optical data.  The weak lensing study (Hoekstra, Franx \& Kuijken~\cite{HFK00}) found three clumps in the mass distribution which all appear to have similar masses. Their central and western clumps seem to correspond to our eastern and western clumps.  However, we do not detect the north-eastern weak lensing clump (see also P. Marshall, these proceedings).

Using the X-ray temperature for the full cluster, $10.4^{+1.7}_{-1.5}$ keV, we can estimate both the velocity dispersion and mass of MS~1054. The implied velocity dispersion is $1289^{+102}_{-96}$ km s$^{-1}$, which agrees well with the optical and weak lensing results.  The virial mass of the cluster can be estimated for $\Omega_m$ = 1 by assuming that the mean density in the virialized region is $\sim$200 times the critical density at the cluster's redshift and that the cluster is isothermal. From the {\itshape Chandra\/} temperature, the virial mass is approximately $6.2^{+1.6}_{-1.3} \times 10^{14} h^{-1} M_{\odot}$ within $r_{200} = 0.76 h^{-1}$ Mpc. This mass is somewhat lower than those derived from weak lensing and the observed velocity dispersion.  Attempts to fit to a $\beta$-model failed because the best fit was obtained for unreasonably large values of $\beta$ and the core radius.  This behavior was also noted in 
Neumann \& Arnaud~\cite{NA00} for the {\itshape ROSAT/HRI\/} data. 

Following line of reasoning similar to that used in Donahue {\it et al}.~\cite{D98}, we determine the expected number density of clusters like MS~1054 in an $\Omega_m$ = 1 universe with initial Gaussian perturbations and compare this with the observed number density for detection in the EMSS.  Conservatively taking the temperature of MS~1054 to be greater than 8.5 keV, we get a minimum cluster mass of $4.5 \times 10^{14} h^{-1} M_{\odot}$. We find a mean virialized mass density of like clusters at z=0.83  $\le 2970 h^2 M_{\odot}$ Mpc$^{-3}$, and a number density $\le 6.6 \times 10^{-12} h^3$ Mpc$^{-3}$.  However, detection of MS~1054 in the EMSS gives a density of like clusters of $\sim 10^{-8} h^3$ Mpc$^{-3}$ (Donahue {\it et al}.~\cite{D98}).  This is greater than our prediction for an $\Omega_m$ = 1 universe with Gaussian perturbations by more than $10^3$.  Although the lower {\itshape Chandra\/} temperature gives an expected number density a factor of 15 higher than that found in Donahue {\it et al}.~\cite{D98}, it does not change their conclusion that $\Omega_m$ = 1 is very unlikely.

Donahue and Voit \cite{DV99} fit for $\Omega_m$ using three cluster samples with different redshift ranges.  They find $\Omega_m \simeq 0.45$ for an open universe and $\Omega_m \simeq 0.27$ for a flat universe.  These results are unaffected by the lower temperature for MS~1054.  In fact, with a temperature of 10 keV, MS~1054 appears to lie closer to the best fit model.

\section{Two High z Quasars: Probing the Hot IGM}

Cosmological numerical simulations suggest that a large fraction of the baryonic matter should reside in intergalactic space in the form of a hot/warm intergalactic medium (IGM) (see, e.g., Fukugita, Hogan \& Peebles~\cite{fhp98}; Cen \& Ostriker~\cite{cos99}; Dav{\' e}, R.\ {\it et al}.~\cite{dco01}). 
 By using numerical simulations Hellsten, Gnedin \& Miralda-Escud\'{e}~\cite{hgm98} explored the idea that a clumpy, highly-ionized IGM concentrated in the gravitational potential caused by dark matter might introduce resonant absorption lines in the X-ray spectrum of a background quasar and denoted them as the ``X-ray Forest'', in analogy to the Ly$\alpha$ forest in the optical/UV bands. Similar ideas were discussed by Perna \& Loeb~\cite{plo98} and Fang \& Canizares~\cite{fca00}.

We used the {\itshape Chandra\/} High Energy Transmission Grating Spectrometer (HETGS) to observe two high-redshift quasars: S5 0836+710 ($z=2.17$) and PKS 2149-306 ($z=2.34$). The HETGS consists of two different grating assemblies, the High Energy Grating (HEG) and the Medium Energy Grating (MEG), and provides nearly constant spectral resolution ($\Delta\lambda = 0.012 \AA$ for HEG and $\Delta\lambda = 0.023 \AA$ for MEG) through the entire bandpass (HEG: $0.8$-$10\ keV$, MEG: $0.4$-$8\ keV$).  Our primary goal is to find if the predicted ``X-ray forest'' is detectable with the {\sl Chandra} HETGS and if not, what are the limits of possible absorbers. 

A complete description of this work is given in Fang {\it et al}.~\cite{F01}. Here we 
summarize the primary results:

\begin{enumerate}

\item The photon index ($\Gamma$) for S5 0836+710 is 1.39, consistent with previous observations with {\sl ROSAT} and {\sl ASCA}. Compared with previous observations, PKS 2149-306 has a rather soft spectrum with $\Gamma=1.255$. Both photon indexes are consistent with the fact that radio-loud quasars (RLQs) have flatter spectra compared with radio-quite quasars (RQQs). We also find excess continuum absorption (above the Galactic value) in S5 0836+710. The flux of S5 0836+710 is higher than in previous {\sl ASCA} observations and puts this source as one of the most luminous objects at high redshift.

\item No significant absorption or emission feature is found in both sources at or above $\pm3\sigma$ level. We put constraints on the possible emitters or absorbers. For example, Fe K emission lines were found in several RQQs, although not as frequently as in RLQs. We put tight upper limits on EW, as low as $\sim 10\ eV$ in both sources. We do not find the emission feature reported by Yaqoob {\it et al}.~\cite{ygn99} in the spectrum of PKS 2149-306 around $5\ keV$. We also examine the possible intervening systems and give upper limits on the absorbers' ion column density. 

\item We discuss the possibility of constraining the properties of the IGM. We develop a method to constrain cosmological parameters based on the X-ray forest method. However, we are not able to set any meaningful constraints based on current data. The X-ray Gunn-Peterson test was also performed and we find that, by combining with the constraints from the CMB Compton $y$-parameter, this method can be applied to constrain a uniform, enriched IGM.

\end{enumerate}

\
\section*{Acknowledgments}
  We thank D. Davis, D. Dewey,D. Huenemoerder, J. Houck, M. Wise, and the rest of the MIT HETG/CXC group. This work was funded by contract SAO SV1-61010, NASA contracts NAS8-39073, NAS8-37716, NAS8-38249, and NAS8-38252, and HST grant GO-6668.  MED was also partially supported by NAG5-3257 and NAG5-6236.  TEJ   acknowledges the support of a NSF fellowship. Support for GLB was provided by NASA through Hubble Fellowship grant HF-01104.01-98A from the Space Telescope Science Institute, which is operated by the Association of Universities for Research in Astronomy, Inc., under NASA contract NAS 6-26555.

\end{document}